\documentclass[a4paper]{article}
\usepackage{latexsym}
\begin{document}

\newtheorem{definition}{Definition}
\newtheorem{theorem}{Theorem}
\newtheorem{proposition}{Proposition}
\newtheorem{remark}{Remark}
\newtheorem{corollary}{Corollary}
\newtheorem{lemma}{Lemma}
\newtheorem{observation}{Observation}
\newtheorem{fact}{Fact}
\newtheorem{example}{Example}

\newcommand{\qed}{\hfill$\Box$\medskip}

\makeatletter
\def\adots{\mathinner{\mkern 1mu\raise \p@ \vbox{\kern 7\p@ \hbox{.}}
\mkern 2mu \raise 4\p@ \hbox{.}\mkern 2mu\raise 7\p@ \hbox{.}\mkern 1mu}}
\makeatother
\newcommand{\ntowers}{NSPACE$\left(2^{2^{\adots^{2}}}\right)$
 }

\title{LIKE Patterns and Complexity} 
\author{Holger Petersen\\
Reinsburgstr.~75\\
70197 Stuttgart\\
Germany} 

%\date{\vspace{-4ex}}
\maketitle

\begin{abstract}
We investigate the expressive power and complexity questions for the LIKE operator in SQL.
\end{abstract}

\section{Introduction}
Regular expressions conveniently support the analysis of software defects involving strings 
stored in a data base and the subsequent selection of test data for checking 
the effectiveness of data cleansing. As an example take a list of
values separated by a special symbol. When manipulating strings of this form, it might happen that 
separator symbols are stored consecutively or strings start with a separator. 
This data corruption
possibly leads to problems when displaying the data or generating export files.

A very restricted variant of regular expressions we will 
consider are patterns for the LIKE operator 
available in SQL (Structured Query Language) \cite{Oracle}. It admits defining patterns
including constants and wild-card symbols representing single letters or arbitrary strings.
Since our investigations are motivated by defect analysis and test data selection, which
by definition may not modify data, we 
assume that new auxiliary columns for holding intermediate values cannot be defined.

Continuing the example given above, we can select corrupt strings in an obvious way.
After data cleansing, the same selections can verify the correctness of the resulting data.
It is known that LIKE pattern matching can define star-free languages only
 \cite[Section~4.2]{BLSS01}.
In Section~\ref{express} we will explore what classes of languages known from the literature
are characterized by LIKE patterns and their boolean combinations.

A more extensive set of operations than those available with the LIKE operator
(including concatenation and closure) is employed
in classical regular expressions studied in Theoretical Computer Science. 
An even more powerful
set of operations is offered by practical regular expressions, which may contain 
back references \cite{Aho90}.

\section{Preliminaries}
For basic definitions related to formal languages, finite automata, and computational complexity we refer to 
\cite{Sipser06}.

The {\em star-free languages} are those regular languages obtained by replacing the star-operator 
with complement in regular expressions. 
Cohen and Brzozowski \cite{CB71} defined a hierarchy of star-free languages according
to the notion of {\em dot-depth}. For an alphabet $\Sigma = \{ a_1, \ldots, a_k\}$
the family $E_0$ consists of the {\em basic} languages $\{ a_1\},\ldots, \{a_k\}, \{\varepsilon\}$ 
(where $\varepsilon$ denotes the empty string).
If $X$ is a family of languages, then we denote by $B(X)$ the boolean closure of $X$ and by
$M(X)$ the closure of $X$ under concatenation. Define the following hierarchy of language families:
\begin{eqnarray*}
B_1 & = & B(E_0) \\
M_n & = & M(B_n) \mbox{ for $n \ge 1$}\\
B_n & = & B(M_{n-1}) \mbox{ for $n \ge 2$}
\end{eqnarray*}
Obviously these families form a hierarchy: 
$$ E_0 \subseteq B_1 \subseteq M_1 \subseteq B_2 \subseteq M_2 \cdots$$
In \cite{CB71} its is shown that the hierarchy is strict up to dot-depth 2 ($B_3$), leaving open whether
the upper levels can be separated. This open problem was resolved in \cite{BK78} by showing that the
hierarchy is strict.
The {\em dot-depth} $d(R)$ for a language $R$ is defined to be $n$ if $R\in B_{n+1} \setminus B_n$. 

The LIKE operator of SQL admits defining patterns in WHERE clauses which can be matched against string valued
 columns . 
Each symbol represents itself except for
certain meta-characters, among which the most important is \% as a wildcard matching zero or more characters. 
Symbol \_  is a substitute for an arbitrary single character. Similarly as further syntactic enhancements (character sets and complements of such sets), the
\_  wildcard can be seen as a (very convenient) shorthand for an enumeration of patterns for every symbol in the alphabet.
If wildcard symbols are required in a pattern, an escape symbol can be declared that enforces 
a literal interpretation of \% and \_.

The more powerful operator SIMILAR TO or the Oracle$^{\tiny\textregistered}$  function REGEXP\_LIKE implement 
general regular expression matching in SQL (the latter even for extended regular expressions). 

The following table compares different notations of the variants
Practical Regular Expressions (PRE),  Classical Regular Expressions (CRE) \cite{Sipser06}, 
Star-Free Expressions (SFE), and LIKE Patterns
\footnote{By 'impl.' we denote the implicit notation of the empty string or concatentaion by juxtaposition of neighboring symbols. $\Sigma$ is not part of the syntax of CRE or 
SFE but a common abbreviation}:

\begin{center}
\begin{tabular}{|l|c|c|c|c|}\hline
   &  PRE & CRE & SFEs & LIKE Patterns\\\hline
letter  $x$     &  $x$ &   $x$   &  $x$  &  $x$ \\\hline
empty string    &   impl. & $\varepsilon$ & $\varepsilon$   &  impl. \\\hline
union              &  $|$     & $\cup$, $+$   & $\cup$, $+$ & n/a \\\hline
concatenation & impl. &  impl., $\circ$, $\cdot$ &  impl., $\circ$, $\cdot$ & impl. \\\hline
closure            & $*$  & $^*$               & n/a & n/a \\\hline
any character  & $.$ & $\Sigma$             & $\Sigma$ & \_  \\\hline
any string        & $.*$  & $\Sigma^*$        & $\overline{\emptyset}$ & \% \\\hline
\end{tabular}
\end{center}
CRE and SFE include a notation for the empty set, which is not relevant for practical purposes and thus does not have a counterpart in 
PRE or LIKE patterns. PRE may include as ``syntactic sugar'' the notations $[\alpha_1\alpha_2\ldots\alpha_n]$ for the set of characters
 $\{\alpha_1, \alpha_2, \ldots, \alpha_n\}$ and  $[\alpha_1-\alpha_n]$ for the range of consecutive characters $\alpha_1$ to $\alpha_n$ 
(this assumes some specific encoding). Notation $[\mbox{\^{}}\alpha_1\alpha_2\ldots\alpha_n]$ and 
 $[\mbox{\^{}}\alpha_1-\alpha_n]$ denote the complements of these sets of characters. Other extensions are the notation $e?$ that denotes zero or one
occurrence of expression $e$ and $e\{n\}$ that denotes exactly $n$ occurrences. None of these operators increases the expressive power of 
regular expressions, but they may lead to significantly shorter expressions than possible with CRE.

One extension of PRE that goes beyond regular languages is the use of back references. The $k$-th subexpression put into parentheses
can be referenced by $\\k$, which matches the same string as matched by the subexpression. 

For CRE the membership problem asks whether the {\em entire} input text match\-es a given pattern. 
In practice we are more interested on one or even all substrings within the input text matching the 
pattern. From the latter set of substrings the answer to the decision problem can easily be derived and
lower bounds above polynomial time carry over (notice that the number of substrings of a text of 
length $n$ is ${{n+1}\choose 2} = \Theta(n^2)$). 
We can enforce a match of a PRE $\alpha$ with the entire input text by enclosing it into ``anchors''
and matching with $\mbox{\^{}}\alpha\$$. Conversely, the CRE $\Sigma^*\alpha\Sigma^*$ simulates the
PRE $\alpha$. We conclude that upper and lower bounds for CRE membership and PRE matching
coincide.

Since LIKE patterns are rather restricted (see Section~\ref{express}) we also consider boolean 
formulas containing LIKE patterns (LIKE expressions) and boolean formulas without negations
(monotone LIKE expressions).
\begin{definition}
A language $L  \subseteq \Sigma^*$  is {\em  LIKE-characterizable} if it is a set of strings 
satisfying a boolean combination of LIKE pattern matching conditions.
 \end{definition}

We summarize known complexity results for some decision problems related to regular expressions:
\begin{center}
\begin{tabular}{|l|c|c|c|}\hline
                                     &  PRE                                                            & CRE & SFE\\\hline
matching,                     &  NP-complete                          &     NL-complete         & P-complete  \\
member                 &  \cite[Thm.~6.2]{Aho90} &         \cite[Thm.~2.2]{JR91}            & \cite[Thm.~1]{Petersen00}  \\\hline
equivalence                &   undecidable               & PSPACE-compl. & $\left.\mbox{\ntowers}\!\!\!\!\right\}{\scriptsize g(n)}$ \\
                                     &      \cite[Thm.~9]{Freydenberger11}   &   \cite[Lem.~2.3]{MS72}   & $g(n) = n$ \cite{SM73} (u.  b.) \\
                                     &                                                               &   & $g(n) = \frac{c\cdot n}{(\log^*n)^2}$ \cite{Fuerer78} (l.  b.) \\\hline
non-                             &      $\in$ ALOGTIME   &     $\in$ ALOGTIME     &  \\
emptiness                     &             &        & see equivalence  \\
                                     &     (see   CRE)                      &     \cite[Intr.]{Petersen00}&  \\\hline  % in P \cite[Lem. 3.1]{Hunt73b}, Petersen00 p. 24 
\end{tabular}
\end{center}

\section{Expressive Power}\label{express}
In this section we briefly discuss the power of LIKE patterns and LIKE expressions in comparison 
to the dot-depth hierarchy as defined in \cite{CB71}.

It is clear that the languages of family $E_0$ can be characterized by LIKE patterns of the form 
$a_i$. Family $B_1$ is incomparable to the languages characterized by LIKE patterns: For an
alphabet $\Sigma$ with $|\Sigma| \ge 2$ the language $L_1 = \{ a_1, \varepsilon\}$ is clearly in $B_1$ 
(a boolean combination of basic languages), but a LIKE pattern characterizing a finite
language can contain different words via \_ only, which allows for words of the same length only.
Thus $L_1$ cannot be characterized by a LIKE pattern. Conversely, the LIKE pattern $00$ defines
the language $L_2 = \{ 00 \}$, which cannot be expressed as a boolean combination of basic languages.
Monotone LIKE expressions can describe all finite languages, but also all co-finite languages. 
Therefore $B_1$ is properly contained in the languages characterized by monotone LIKE expressions 
(separation by $L_2$).

Every language $R$ in family $B_2$ can be denoted in the form
$$ \bigcup_{k=1}^{\ell} \left(\left[\bigcap_{i=1}^{m(k)} 
  \overline{ w_0^{k,i}\Sigma^* w_1^{k,i}\Sigma^*\cdots \Sigma^* w_{s(k,i)}^{k,i}}\right]
 \cap 
  \left[\bigcap_{j=1}^{n(k)} 
   u_0^{k,j}\Sigma^* u_1^{k,j}\Sigma^*\cdots \Sigma^* u_{t(k,j)}^{k,j}\right]\right)
$$
with $w_p^{k,i}$, $u_q^{k,j}$ words and $m(k)$, $n(k)$, $\ell$, $s(k,i)$, $t(k,j)$ 
non-negative integers \cite[Lemma~2.8]{CB71}.
This representation translates directly to a LIKE expression. Given a LIKE expression,
every pattern containing wildcards \_ can be replaced by an enumeration of patterns
substituting the alphabet symbols for wildcards. All negations can be moved to the 
LIKE operators applying De Morgan's laws. The resulting expression characterizes 
a set in $B_2$.

We are thus led to the following observation:
\begin{observation}
The class of LIKE-characterizable languages coincides with the class of languages of dot-depth 1.
\end{observation}
An example of a star-free language shown to be of dot-depth 2 (and therefore not LIKE-characterizable)
is $(0+1+2)^*02^*$ from \cite[LEMMA~2.9]{CB71}.

Finally we sketch why monotone LIKE expressions are weaker than general LIKE expressions. 
We claim that monotone LIKE expressions cannot express that strings are formed over a
proper subset $\Sigma'$ of the underlying alphabet $\Sigma$  
(which we assume to contain at least two symbols). 
Suppose a monotone LIKE expression $e$ can express this restriction. Choose a string $w$ over
$\Sigma'$ which is longer than $e$. Then $w$ matches $e$ and at least one symbol of $w$ matches
wildcards only. This symbol can be substituted by a symbol from $\Sigma\setminus\Sigma'$. 
The resulting string still matches $e$, contradicting the assumption.  

\section{Computational Complexity}
We first introduce a syntactical transformation of patterns that will simplify the 
subsequent algorithms.
\begin{definition}\label{defnormalize}
A LIKE pattern is called {\em normalized}, if it contains none of the substrings
\%\_ and \%\%.
\end{definition}

Consider an arbitrary string $w \in \{\% , \_\}^* $ consisting of wildcards. If $w$ 
matches a string over the base alphabet, then a string $w'$ containing the same number of the symbol \_ and 
a trailing \% if and only if $w$ contains \% matches as well. Since $w'$ is normalized we obtain:
\begin{proposition}\label{propnormalize}
For every LIKE pattern there is an equivalent normalized LIKE pattern.
\end{proposition}
Normalization cannot in general identify equivalent patterns. As an example take the
patterns $\%01\%$ and $\%0\%1\%$ over the binary alphabet $\{ 0, 1\}$. Obviously, any string matching the 
first pattern matches the second. But the converse is also true, because there is a left-most $1$ between
the two constants of the pattern (including the $1$) and it is preceded by a $0$. 
Over the alphabet $\{ 0, 1, 2\}$, the patterns are separated by $021$.

\begin{lemma}\label{lemmanormalize}
LIKE patterns can be normalized in deterministic logarithmic space.
\end{lemma}
{\bf Proof.}
Any input can be written as $x_0w_0\cdots x_nw_nx_{n+1}$ where 
$w_0\cdots w_n \in \{\% , \_\}^* $ and $x_0\cdots x_{n+1} \in \Sigma^* $
for the underlying alphabet $\Sigma$.

A deterministic Turing machine $M$ scans the input and directly outputs any symbol from $\Sigma$.
For every string $w_i$ of consecutive wildcards, the number $m$ of occurrences of \_ is counted 
and a flag is maintained indicating the presence of \%. At the end of $w_i$, machine $M$ outputs
$m$ symbols \_ and an optional \% if the flag is set.

Since $M$ has to store counters bounded by the input length, it can do so in logarithmic
space if the counters are encoded in binary notation.
\qed 

\begin{theorem}\label{membershipLIKEdlogspace}
Matching with a LIKE pattern can be done in deterministic logarithmic space.
\end{theorem}
{\bf Proof.}
If the pattern contains no \%, in a single scan the constant symbols in the pattern are 
compared and for every \_ in the pattern a symbol in the text is skipped.
 
By Lemma~\ref{lemmanormalize} we can assume that any LIKE-pattern containing \%
has the form $p = a_1\%a_2\%\cdots \%a_n$ where $a_i \in (\Sigma\cup\{ \_\})^*$.
We first argue that a greedy matching strategy suffices for checking whether $p$ matches a text
$t$. Suppose in a given matching $i$ is minimal with the property that $a_i$ could be
matched further to the start of the text (but after $a_{i-1}$). Then a new match can be
obtained by moving $a_i$ to the first occurrence. Carrying out this operation for all $a_i$
leads to a greedy matching.

For every $a_i$ a left-most match can be determined by comparing the constant part and
shifting the position in the text if a mis-match occurs. Once an $a_i$ has been matched,
it is not necessary to reconsider it by the argument above.

In logarithmic space pointers into pattern and text can be stored and by scanning $p$ and 
$t$ in parallel a greedy matching can be determined. \qed 

We have the following (weaker) lower bound for the membership problem: 
\begin{theorem}
Matching with a LIKE pattern cannot be done by constant-depth, polynomial-size, unbounded
fan-in circuits (it is not in AC$^0$).
\end{theorem}
{\bf Proof.}
Recall from \cite{FSS84} that the majority predicate on $n$ binary variables is $1$ if
and only if more than half of the input values are 1. We map a given input $x$ for the 
majority predicate to the pattern $\%(1\%)^{\lceil (|x|+1)/2\rceil}$. String $x$ matches the
pattern only if $x$ contains at least $\lceil (|x|+1)/2\rceil > |x|/2$ symbols $1$, which 
is majority. By the result \cite[Theorem 4.3]{FSS84} this predicate is not in AC$^0$. 
\qed

Since the evaluation of boolean formulas is possible in logarithmic space, we
obtain from Theorem~\ref{membershipLIKEdlogspace}:
\begin{corollary}\label{membershipLIKEexprdlogspace}
Matching with a LIKE expression can be done in deterministic logarithmic space.
\end{corollary}

Considering equivalence of LIKE patterns, a test using syntactical properties 
alone seems to be impossible because of the example given above. 

Based on Theorem~\ref{membershipLIKEdlogspace} we can obtain the following upper bound:
\begin{corollary}
Inequivalence of LIKE patterns is in nondeterministic logarithmic space.
\end{corollary}
{\bf Proof.}
Guess a separating text symbol by symbol and match with the given patterns in logarithmic space.
 \qed

\begin{theorem}\label{nonempmonLIKE}
Nonemptiness of monotone LIKE-expressions is complete in NP.
\end{theorem}
{\bf Proof.}
For membership in NP consider a string $w$ matching a given expression $e$. 
We claim that there is no loss of generality in assuming $|w| \le |e|$.
We fix a matching of $w$ by $e$. 
For every OR in expression $e$ there has to be at least one sub-expression matching 
$w$. We delete the other sub-expression and continue this process
until there is no OR left obtaining $e'$. Clearly $|e'| \le |e|$. Now we mark 
every symbol of $w$ matched by a constant or \_. At most $|e'|$ symbols of $w$ will
thus be marked and the others have to be matched by \%. Deleting these symbols
yields a string $w'$ matching $e$ with $|w'| \le |e'| \le |e|$. The NP algorithm
simply consists in guessing a string $w$ with $|w|\le |e|$, writing it onto
the work tape, and checking membership according to Corollary~\ref{membershipLIKEexprdlogspace}.

For hardness we reduce the satisfiability problem of boolean formulas in 3-CNF (3SAT) to the 
nonemptiness problem. It is well-known that 3SAT is complete in NP \cite{Sipser06}. 
Let 
$$F = (\alpha_1 \vee \beta_1 \vee \gamma_1) \wedge \cdots \wedge (\alpha_m \vee \beta_m \vee \gamma_m)$$ 
be a formula in CNF over variables $x_1, \ldots, x_n$. The idea is to enumerate
all satisfied literals in a string that matches a monotone LIKE-expression.
We form a set of LIKE patterns over the alphabet $\{x_1, \ldots, x_n, \bar x_1, \ldots, \bar x_n\}$
that are joined by AND:
\begin{itemize}
\item $\_^n$ (there are exactly $n$ literals).
\item For $1 \le i \le n$ an OR of the patterns $x_i$ and $\bar x_i$ (for every variable 
  at least one literal is true).
\item For every clause $\alpha_k \vee \beta_k \vee \gamma_k$ an OR of the patterns
  $\alpha_k$, $\beta_k$, and $\gamma_k$ (at least one literal is true in every clause).
\end{itemize}
Suppose that $F$ is satisfied by some assignment of boolean values to 
$x_1, \ldots, x_n$. Concatenate the satisfied literal for each variable to form a string
to be matched. This string clearly matches all patterns defined above. 
Conversely, if a string matches all patterns it contains at least one literal per variable by the
second item. The length restriction to $n$ symbols implies that exactly one literal per variable
is included. These literals define a truth assignment in the obvious way and by the third item
every clause is satisfied by this assignment. 
\qed

\begin{lemma}\label{lemmavalidTM}
For a deterministic Turing machine $M$ with input $w$ and space bound $s(|w|)$, 
a LIKE-expression $e$ with the following properties can be constructed:
\begin{enumerate}
\item All LIKE conditions are negative.
\item The LIKE-expression $e$ is of size $O(s^2(|w|)$.
\item If $M$ accepts $w$ within space $s(|w|)$, there is a single string matching $e$.
\item If $M$ does not accept $w$ within space $s(|w|)$, the language described by $e$ is empty.
\end{enumerate}
\end{lemma}
{\bf Proof.}
Without loss of generality we assume that $M$ accepts with a blank tape and the tape head
on the left-most tape cell. We denote the input length by $n = |w|$.

In order to simplify the presentation we first use arbitrary LIKE conditions. We encode 
a computation of $M$ as a sequence of {\em configurations} over the alphabet $\Gamma \cup Q$ 
(tape alphabet and set of states). A configuration $uqv$ encodes the tape inscription $uv$,
current state $q$ and head position on the first symbol of $v$. A computation consisting of
$k$ steps is encoded as $\#c_0\#c_1\#\cdots\#c_k\#$. Configuration $c_0$ is $q_0w$ followed
by $s(n)-n$ blanks and for $i \ge 1$ configuration $c_{i-1}$ yields $c_{i}$ by $M$'s transition
function. We therefore identify the following patterns:
\begin{enumerate}
\item\label{start_conf} $\#c_0\#\%$ (start configuration).
\item\label{acc_conf} $\%\#c_{\mbox{accept}}\#$ (accepting configuration).
\item For every $\delta(q_i, b) = (q_j, c, L)$ negative patterns $aq_ib\_^{s(n)}def$ with 
  $def \neq q_jac$.
\item For every $\delta(q_i, b) = (q_j, c, R)$ negative patterns $aq_ib\_^{s(n)}def$ with 
  $def \neq acq_j$.
\item Negative patterns $abc\_^{s(n)}d$ with $a, b, c \in \Gamma \cup \{\#\}$ and $b \neq d$ 
 (portions of the tape not affected by the computation).
\end{enumerate}
For each of the patterns in item~\ref{start_conf} and \ref{acc_conf} we can substitute  
$(s(n)+2)(|\Gamma| + |Q|)$ equivalent negative patterns that exclude all but one symbol from
$\Gamma \cup Q \cup \{ \# \}$ at position $i$ with $1 \le i \le s(n)+2$ from the start resp.\ 
end of the string.
\qed

\begin{lemma}\label{lemmaupperineq}
Inequivalence of LIKE-expressions can be decided nondeterministically in linear space.
\end{lemma}
{\bf Proof.}
For two given expressions guess a string symbol by symbol and mark in every pattern the 
positions reachable by matching the guessed string. When a separating string has been found, both 
expressions are evaluated and it is checked that exactly one of the expressions matches.
\qed

The previous lemmas can be summarized in the following way:
\begin{theorem}
Equivalence of monotone as well as of arbitrary LIKE-expressions is complete in PSPACE.
\end{theorem}

\section{Discussion}
We investigated the expressive power and computational complexity of the LIKE operator.
For the more powerful monotone and general LIKE expressions we classified the 
complexity of nonemptiness and equivalence. In case of membership we could 
establish the upper bound  L (deterministic logarithmic space). This is believed to
be of lower complexity than the general membership problem for CRE, which is complete
in NL \cite{JR91}. 
Membership for a single LIKE pattern is not decidable by the highly parallel AC$^0$ circuits.
It remains open, what the exact complexity of the latter problem and inequivalence is.

\section*{Acknowledgement}
Many thanks to Manfred Kuf{}leitner for information about star-free languages.

%\end{document}

\end{document}